\documentstyle[aps,12pt]{revtex}
\input{psfig.sty}

\protect
\begin{document}
\title{2+ P-SAT: Relation of Typical-Case Complexity to the Nature of the
Phase Transition}

\author{
R\'emi Monasson \cite{rm}, Riccardo Zecchina \cite{rz},
Scott Kirkpatrick \cite{sk}, Bart Selman \cite{bs}, Lidror Troyansky \cite{lt}
}

\address{
\cite{rm} Laboratoire de Physique Th\'eorique de l'ENS, Paris, France\\
\cite{rz} International Centre for Theoretical Physics, Trieste, Italy\\
\cite{sk} IBM, Thomas J. Watson Research Center,Yorktown Heights, NY \\
\cite{bs} Computer Science Dept., Cornell University, Ithaca, NY \\
\cite{lt} Computer Science Dept., Hebrew University, Jerusalem, Israel\\
}


\maketitle

\begin{abstract}
\thispagestyle{empty}
Heuristic methods for solution of problems in the NP-Complete class of 
decision problems often reach exact solutions, but fail badly at 
"phase boundaries," across which the decision to be reached changes 
from almost always having one value to almost always having a different 
value.  We report an analytic solution and experimental investigations of 
the phase transition that occurs in the limit of very large problems 
in K-SAT.  The nature of its "random first-order" phase transition, seen 
at values of K large enough to make the computational cost of solving 
typical instances increase exponentially with problem size, suggests a 
mechanism for the cost increase.  There has been evidence for features 
like the "backbone" of frozen inputs which characterizes the UNSAT phase 
in K-SAT in the study of models of disordered materials, but this feature 
and this transition are uniquely accessible to analysis in K-SAT.  The 
random first-order transition combines properties of the 1st order 
(discontinuous onset of order) and 2nd order (with power law scaling, 
e.g. of the width of the critical region in a finite system) transitions 
known in the physics of pure solids.  Such transitions should occur 
in other combinatoric problems in the large N limit. Finally, improved 
search heuristics may be developed when a "backbone" is known to exist.

\end{abstract}

\thispagestyle{empty}
\newpage
\setcounter{page}{1}

Constraint satisfaction, the automated search for a configuration of a
complex system which satisfies a set of rules or inequalities, is
often difficult, and occurs widely in practice.  The simplest example
of a constraint satisfaction problem, K-SAT\cite{Hayes,exact}, is
commonly used as a testbed for heuristic algorithms intended for wider
use and was the first problem proved to be in the complexity class
NP-Complete\cite{Cook,NPC}, in which the worst case instances are believed to
always require computing effort exponential in $N$, the number of input
degrees of freedom.  In random K-SAT, the system parameters are a string of
$N$ bits, and the rules to be satisfied are a set of $M$ clauses.  If
the string consists of bits $\{x_i=0,1\}_{i=1,\ldots,N}$, we construct
an instance of K-SAT by first randomly choosing $K$ distinct
possible indices $i$ and then, for each of them, a literal $z_i$ (i.e.
the corresponding $x_i$ or its negation $\bar x_i$ with equal
probability).  A clause $C$ is defined as the logical OR of the $K$
literals chosen. Next, we repeat this process to obtain $M$
independently chosen clauses $\{C_\ell\}_{\ell=1,\ldots,M}$ and ask
for all of them to be true at the same time, i.e. we take the logical
AND of the $M$ clauses.  This gives a formula $F$, which may be
written as
\begin{equation}
F=\bigwedge_{\ell =1}^M C_\ell =\bigwedge_{\ell = 1}^M\;\;\left(
\bigvee_{i =1}^K z_i ^{(\ell )}\right) \;\;\;,
\label{Fcnf}
\end{equation}
where $\bigwedge$ and $\bigvee$ stand for the logical AND and OR
operations respectively.  An assignment of the $\{x_i\}$'s satisfying
all clauses is a solution of the K--SAT problem. If no such
assignment exists, $F$ is unsatisfiable.  The formulae $F$ constructed
at random keeping the ratio $\alpha = M/N$ constant as $M,N \to
\infty$ provide a natural ensemble of test problems, with $\alpha$
characterizing whether the $F$ are typically under- or
over-constrained.

The value of K is important.  2--SAT can be solved by a linear time
algorithm\cite{algo}. There is a critical value of $\alpha$,
$\alpha_c(2) = 1$, below which the likelihood of an $F$ being UNSAT
vanishes in the limit $N \to \infty$, and above which it goes to 1.
For $K \ge 3$, K--SAT is NP-Complete and rigorous results are few.
Computer experiments \cite{KirkSel,NIPS} on K-SAT for $K=2,3$ and
higher have located the phase transition and provided critical
exponents for the sharpening of the critical region which occurs with
increasing sample size $N$.

The technique used, finite-size scaling, will be
discussed in more detail below.  It has recently been
proven\cite{Friedgut} that certain monotonic properties (such as SAT)
in combinatorial ensembles do have sharp thresholds.  ``Sharp'' means
in our case that for any $ \alpha < \alpha_c(K,N) $, the probability
that a formula in K-SAT can be satisfied goes to 1 as $N \to
\infty $ , while for any $ \alpha > \alpha_c(K,N)$ this probability
tends to 0.  While Friedgut's result leaves open the question of whether
$\alpha_c(K,N)$ has a limit
as $N \to \infty$, experiment suggests that this is the case for K-SAT.

One reason for the recent interest in the threshold is the growing recognition
that ``easy-hard-easy'' patterns of computational cost are
characteristic of heuristic algorithms developed to solve problems
whose cost in the worst case increases exponentially with problem size
N, and that the hardest instances occur near \cite{note1} phase boundaries like
$\alpha_c$\cite{mitchell:hard,AI,SelKirk}.

There is a strong analogy between these problems and the properties
of disordered materials, alloys or even glasses, studied by constructing
models whose energy function expresses the constraints\cite{SimAnn}.
Strongly disordered models with conflicting
interactions similar to the randomly negated literals in K--SAT are known as "spin
glasses"\cite{kro}.  Fu and Anderson\cite{FuAnd} first conjectured that spin glasses
are the models underlying NP-Complete decision problems and vice-versa.  They cite
the example of weighted graph partitioning, which is equivalent to an Ising or Potts
spin glass.

A technique of calculating expectation values of observables in random many-parameter
systems, called the replica method\cite{MPV} predicts that
ordering of a new type is possible in the presence of
microscopic randomness.  In the replica method, the calculation of the average over the
disorder leads to an effective energy or cost function which describes
many identical copies of one instance of the model system
with the dynamical variables (usually Ising spins taking on values +1 or -1)
in different replicas coupled by some non-linear function.  The onset of ordering can
be identified by the fact that a single stable state occurs in
multiple replicas.  This corresponds to a physical ground state that is
highly irregular in structure due to the randomness of the problem. A
more subtle kind of ordering (called Replica Symmetry-Breaking) occurs
when distinct stable states of this sort are found in different
subsets of the replicas, signalling that there may be infinitely many distinct ground
states with energies infinitesimally close to the true ground state.
The extent of or absence of this new kind of order can be quantified by an
``order parameter,'' which in
general emerges from the replica formulation. We describe both types of result
in sections I and II. Details of the calculations for K-SAT are given in
section II. While the replica procedures are not rigorous, certain
steps can be proven, and some results have been verified by other means.  We 
discuss these issues in section II.

Spin glass models with realistic connectivity are difficult to explore
experimentally, either on real substances or by computer simulation of models
in thermal equilibrium at some finite temperature.
Experimental study of the ``easy-hard-easy'' phase transitions in
combinatorics is more tractable. Although these are spin glass models, the
properties of interest are
ground state properties, and a large body of model-specific heuristics exists,
which gives powerful means of exploring these ground states.  We have previously
applied replica methods and determined characteristics of the 3-SAT
transition\cite{MZI,MZII,MZIII}. In section III, we report additional results
which provide new insights, hopefully of use to both fields.
Note, however, that the methods of statistical
physics predict the most probable, or typical, behavior of a system
with many degrees of freedom, so we shall be describing the typical
complexity of K--SAT, not its worst case.

\section{Mixtures of $k=2$ and 3: Overview of Results}
In order to understand what occurs between $K=2$ and $K=3$, we have
studied\cite{physcomp96} formulae containing mixtures of 2- and 3-clauses:
consider a random formula with M clauses, of which $(1-p)M$ contain two
literals and $pM$ contain 3 literals, with
$0\le p\le 1$. This ``$2+p$--SAT'' model
smoothly interpolates between 2--SAT $(p=0)$ and 3--SAT $(p=1)$.
The problem is NP--complete, since any instance of the model for $p>0$
contains a sub-formula of 3-clauses. But our interest here is in the
complexity of ``typical'' problem instances.

We seek $\alpha_c (2+p)$, the threshold ratio $M/N$ of the above model at
fixed $p$. We know $\alpha_c (2)=1$ and $\alpha_c (3)\simeq
4.27$.  $F$ cannot be almost always satisfied if the number of
2--clauses (respectively 3--clauses) exceeds $N$ (resp. $\alpha _c (3) N$).
As a consequence, the critical ratio must be bounded by
$\alpha_c(2+p)\le\hbox{min} \left( \frac{1}{1-p} ,
\frac{\alpha _c (3)}{p} \right)$.

The $2+p$--SAT model can be mapped onto a diluted spin
glass model with $N$ spins $S_i$:
$S_i=1$ if the Boolean variable $x_i$ is true, $S_i=-1$ if
$x_i$ is false. Then, to any configuration is associated an energy
$E$, or cost-function, equal to the number of clauses violated.
Random couplings between the spins are induced by the
clauses. {\em The most important result of the replica approach\cite{MZII}
is the emergence, in the large $N,M$ limit and at fixed $p$ and $\alpha$,
of order parameters describing the statistics of optimal assignments, which
minimize the number of violated clauses.} In this section, we give an
overview of the results from statistical mechanics. The next section gives
a more detailed description of the analysis.

Consider an instance of the
$2+p$-SAT problem.  We use the ${\cal N} _{GS}$ ground state
configurations to define
\begin{equation}
m_i = \frac{1}{{\cal N} _{GS}} \sum _{g=1} ^{{\cal N} _{GS}}S_i ^g
\end{equation}
the average value of spin $S_i$ over all optimal configurations. Clearly, $m_i$
ranges from $-1$ to $+1$ and $m_i =-1$ (respectively $+1$) means that
the corresponding Boolean variable $x_i$ is always false (resp. true)
in all ground states.  The distribution $P(m)$ of all $m_i$ gives the
microscopic structure of the ground states.  The accumulation of
magnetizations $m$ around $\pm 1$ represents a ``backbone'' of
almost
completely constrained variables, whose logical values cannot vary
from solution to solution, while the center of the distribution
$P(m\simeq 0)$ describes weakly constrained variables.
The threshold $\alpha _c$
will coincide with the appearance of an extensive backbone density of fully
constrained variables $x_i$, with a finite
probability weight at $m=\pm 1$.
A simple argument shows that the backbone must vanish when
$\alpha < \alpha_c$.  Consider adding
one clause to a SAT formula found below $\alpha_c$.  If there is a finite
fraction of backbone spins, there will be a finite probability that the
added clause creates an UNSAT formula, which cannot occur.

For $\alpha < \alpha _c$, the solution exhibits a simple symmetry property,
usually referred to as Replica Symmetry (RS), which leads to an order
parameter which is precisely the magnetization distribution $P(m)$ defined
above.
An essential qualitative difference
between 2-SAT and 3-SAT is the way the order parameter $P(m)$ changes
at the threshold. This discrepancy can be
seen in the fraction $f(K , \alpha )$ of Boolean variables which become fully
constrained, at and above the threshold. As said above, $f (K, \alpha
)$ is identically null below the threshold. For 2-SAT, $f(2 ,\alpha )$
becomes strictly positive above $\alpha _c=1$ and is continuous at the
transition~: $f(2 , 1^- ) = f(2, 1^+ ) = 0$. On the contrary, $f(3 ,
\alpha )$ displays a discontinuous change at the threshold~:
$f(3, \alpha _c ^- ) =0$ and $f(3, \alpha _c ^+ ) = f_c(3) > 0$.

While for the continuous transitions, the exact value of the threshold
can be derived within the RS scheme, for the discontinuous case the RS
prediction gives only upper bounds. The exact value of the threshold
can be predicted only by a proper choice of the order parameter at the
transition point, i.e. by a more general symmetry breaking scheme, a
problem which is still open. However, the predictions of the RS
equations, such as the number of solutions, remain valid up to
$\alpha_c$, and the RS prediction for the nature of the threshold
should be qualitatively correct.

For the mixed $2+p$--SAT model, the key issue is therefore to
understand how a discontinuous 3--SAT-like transition may
appear when increasing $p$ from zero up to one and how it
affects the computational cost of finding solutions near threshold.
Applying the method of ref.\cite{MZII}, we find for $p<p_0$
($p_0=0.41$), there is a {\bf continuous} SAT/UNSAT transition at
$\alpha_c (2+p) =\frac{1}{1-p}$ .  This has recently been verified by
rigorous analysis up to $p = 0.4$\cite{Achlioptas}.  The RS theory
appears to be correct for $\alpha < \alpha_c (2+p)$, and thus gives
both the critical ratio and the typical number of solutions, as in the
$K=2$ case.  The SAT/UNSAT transition should coincide with a replica
symmetry breaking transition, as discussed in \cite{MZII}.  So, for
$p<p_0$, the model shares the characteristics of random $2$--SAT.

For $p>p_0$, the transition becomes {\bf discontinuous} and the RS
transition gives an upper bound for the true $\alpha_c(2+p)$.  The RS
theory correctly predicts a discontinuous appearance of a finite
fraction of fully constrained variables which jumps from $0$ to $f_c$
when crossing the threshold $\alpha _c (2+p)$. However, both values of
$f_c (2+p )$ and $\alpha _c $ are slightly overestimated, e.g. for $p=1$,
$\alpha _c ^{RS}(3)\simeq 4.60 $ and $f_c ^{RS} (3)\simeq 0.6 $
whereas experiments give $\alpha _c (3) \simeq 4.27 $ and $f_c (3)
\sim 0.5 $.  A replica symmetry breaking theory will be necessary to
predict these quantities.  For $p>p_0$, the random 2+p--SAT problem
shares the characteristics of random $3$--SAT.

This transition differs from phase transitions in most ordered materials in
that the ground state is highly degenerate at $\alpha_c$.
The entropy, i.e. the logarithm (base 2) of the typical number of
optimal solutions, can be computed exactly within the RS scheme for
any $p$ and $\alpha < \alpha_c(2+p)$. The entropy at the transition
point decreases as a function of $p$, from $0.56$ for $p=0$ to $0.13$
for $p=1$, as plotted in Fig. 1.

\section{Statistical mechanics analysis of the 2+P--SAT model}

In this section we describe the analytical
calculation of the typical ground state properties of
the 2+P--SAT model using the replica method.
A brief discussion concerning rigorous results and 
prospects for making the replica results rigorous
is also included.

\subsection{The energy-cost function}
The logical variables $x_i$ can be
represented by $N$ binary variables $S_i$, called spins, through the
one-to-one mapping $S_i=-1$ (respectively $+1$) if $x_i$ is false
(resp. true). We then encode the random clauses into a $M\times N$
matrix $C_{\ell i}$ in the following way~: $C_{\ell i}=-1$
(respectively $+1$) if the clause $C_\ell $ includes $\overline{x_i }$
(resp. $x_i$), $C_{\ell i}=0$ otherwise. Note that
$\sum _{i=1}^N C_{\ell i}
S_i$ gives the the net number of literals satisfying clause $\ell$. Consider now the
cost-function $E[{\bf C } , {\bf S }]$ defined as the number of
clauses that are not satisfied by the logical assignment corresponding
to configuration ${\bf S}$.

\begin{equation}
E[{\bf C},{\bf S}]=  \sum _{\ell =1}^{(1-p)M} \delta
\left[ \sum _{i=1}^N C_{\ell i}
S_i ; -2 \right] + \sum _{\ell =(1-p)M+1}^{M} \delta \left[
\sum _{i=1}^N C_{\ell i} S_i ; -3 \right] \quad ,
\label{cost}
\end{equation}
where $\delta [.;.]$ denotes the Kronecker function, which is 1 if its 
arguments are equal, zero otherwise.
The minimum (or ground state) $E[{\bf C}]$ of $E[{\bf C } ,
{\bf S }]$, the lowest number of violated clauses that can be
achieved by the best possible logical assignment \cite{MZII}, is a
random variable which becomes totally concentrated around its mean
value $\ll E[{\bf C}] \gg$ in the large size limit \cite{selfave}. The
latter is accessible through the knowledge of the averaged logarithm
of the generating function
\begin{equation}
Z[{\bf C } ]= \sum _{\bf S} \exp \left( - E[{\bf C } , {\bf S }] / T \right)
\label{partfunc}
\end{equation}
since $\ll E[{\bf C } ] \gg = - T \ll \log Z[{\bf C } ] \gg + O(T^2)$
when the auxiliary parameter $T$ is eventually sent to zero.
Since $\ll E[{\bf C } ] \gg = 0$ in the SAT
region and is positive in the UNSAT phase, calculating $\ll E[{\bf C}] \gg$
locates $\alpha _c(K)$.

\subsection{The average over the disorder}

The calculation of the average value of the logarithm of $Z$ from Eq.
(\ref{partfunc}) is an awkward one. To circumvent this difficulty, we
compute the $n^{th}$ moment of $Z$ for integer-valued $n$ and perform
an analytical continuation to real $n$ to exploit the identity
$\ll Z[{\bf C } ] ^ n  \gg = 1 + n  \ll \log  Z[{\bf C } ] \gg +
O(n^2)$. The $n^{th}$ moment of $Z$ is obtained by replicating $n$
times the sum over the spin configuration ${\bf S}$ and averaging over
the clause distribution \cite{MZII}
\begin{equation}
\ll Z[{\bf C } ] ^ n  \gg = \sum _{{\bf S}^1 , {\bf S}^2 , \ldots ,
{\bf S}^n} \ll \exp \left( -  \sum _{a=1} ^n E[{\bf C } ,
{\bf S }^a] / T \right) \gg \quad .
\label{nthmoment}
\end{equation}
The average over the clauses can be performed because their 
probability distributions are uncorrelated. We obtain
\begin{equation}
\ll Z[{\bf C } ] ^ n  \gg  = \sum _{{\bf S}^1 , {\bf S}^2 , \ldots ,
{\bf S}^n} \left( \zeta _2 [ S^a ] \right) ^{(1-p) M}
\    \left( \zeta _3 [ S^a ] \right) ^{p M}
\qquad ,
\end{equation}
where each factor is defined by ($K=2,3$)
\begin{equation}
\zeta _K [ S^a ] = \ll \exp \left( -\frac{1}{T} \sum _{a=1} ^n
\delta \left[ \sum _{i=1}^N C_i S_i^a ; -K \right] \right) \gg
\qquad . \label{appe}
\end{equation}
We stress that $\ll .\gg$ now
denotes the unbiased average over the set of $2^K {N \choose K}$
vectors of $N$ components $C_i =0 ,\pm 1$ and of squared norm equal to $K$.

Resorting to the identity,
\begin{equation}
\delta \left[ \sum _{i=1}^N C_i S_i^a ; -K \right] =
\prod _{i / C_i \ne 0} \delta \left[ S_i^a ; -C _i \right] \qquad,
\end{equation}
one may carry out the average over disorder in eq.(\ref{appe}) to obtain
\begin{equation}
\zeta _K [ S^a ] = \frac 1{2^K} \sum _{C_1 ,\ldots , C_K =\pm 1} \frac 1{N^K}
\sum _{i_1 ,\ldots , i_K =1} ^N \exp \left\{ -\frac{1}{T} \sum _{a=1} ^n
\prod _{\ell =1}^K \delta \left[ S_{i_\ell} ^a ; -C _\ell\right] \right\}
\quad ,
\label{zeta1}
\end{equation}
to the largest order in $N$.

It is crucial to remark that $\zeta _K [ S^a ] $ in (\ref{zeta1})
depends on the $n\times N$ spins only through some
$2^n$ quantities $c({\bf \sigma} )$ labelled by the vectors ${\bf \sigma}$
with $n$ binary components; $c({\bf \sigma} )$ equals the number (divided
by $N$) of labels $i$ such that $S_i^a=\sigma ^a$, $\forall a=1,\ldots
,n$ \cite{eilat}. Indeed,
one can rewrite $\zeta _K [ S^a ] =\zeta _K [ c ]$ with
\begin{equation}
\zeta _K [ c ] = \frac 1{2^K} \sum _{C_1 ,\ldots , C_K =\pm 1}
\sum _{\vec \sigma _1 , \ldots , \vec \sigma _K}
c (- C_1 \;\vec \sigma _1  ) \ldots
c (- C_K\; \vec \sigma _K )\exp \left\{
-\frac{1}{T} \sum _{a=1} ^n \prod _{\ell =1}^K \delta \left[ \sigma _\ell
 ^a ; 1 \right] \right\}
\quad .
\end{equation}
Notice that $c (\vec \sigma ) = c (-\vec \sigma ) $ due to the
even distribution of the disorder $C$.

Introducing the {\em effective energy} function,
\begin{eqnarray}
E_{eff} [c ] &=& \alpha\; (1-p) \ \ln \left[ \sum
_{\vec \sigma , \vec \tau} c (\vec \sigma ) c (\vec \tau ) \
\exp \left( - \frac{1}{T} \sum _{a=1} ^n \delta [ \sigma _a ; 1] \delta [
\tau _a ; 1] \right) \right] \nonumber \\ && + \alpha
\;p \ \ln \left[ \sum _{\vec \sigma , \vec \tau , \vec \omega } c
(\vec \sigma ) c (\vec \tau ) c (\vec \omega ) \ \exp \left( -
\frac{1}{T} \sum _{a=1} ^n \delta [ \sigma _a ; 1] \delta [\tau _a ; 1]
\delta [\omega _a ; 1] \right) \right] \qquad ,
\label{expre}
\end{eqnarray}
we may rewrite the $n^{th}$ moment of the generating function $Z$
(\ref{nthmoment}) as
\begin{equation}
\ll Z ^ n  \gg = \int \prod _{\vec \sigma} dc (\vec \sigma
) e^{-E_{eff} [c ] } \; \sum _{{\bf S}^1 , {\bf S}^2 , \ldots ,
{\bf S}^n} \prod _{\vec \sigma} \delta \left( c (\vec \sigma )
-\frac 1N \sum _{i=1}^N \prod_{a=1}^n \delta[ S_i^a ; \sigma ^a] \right)\quad .
\label{nthmoment1}
\end{equation}
The sum over the spins in the last term of the above equation can
be computed, and  gives rise to a combinatorial factor
\begin{equation}
\frac{ N!} {\prod _{\vec \sigma} ( N c (\vec \sigma ) )!}
= \exp \left( -N \sum _{\vec \sigma} c (\vec \sigma ) \ln c (\vec \sigma )
\right) \quad,
\end{equation}
to the leading exponential order in $N$. As a consequence,
the $n^{th}$ moment of $Z$ using the Laplace method is
$\ll Z ^n \gg \simeq \exp ( N \; F_{max})$ where $F_{max}$ is the
maximum over all possible $c$s of the functional \cite{MZII}
\begin{equation}
F[c ] = - \sum _{\vec \sigma} c({\vec \sigma}) \log c({\vec
\sigma}) - T E_{eff}[c ] \; \; \; \qquad ,
\label{eqfunc}
\end{equation}
with the constraint
\begin{equation}
\sum _{\vec \sigma} c({\vec \sigma}) = 1.
\end{equation}

\subsection{The replica symmetric theory}
The optimisation conditions over $F[ c]$ provide $2^n$ coupled
equations for the $c$s. Notice that $F$ is a symmetric functional,
invariant under any permutation of the replicas $a$. A maximum
may thus be sought in the so-called replica symmetric (RS) subspace of
dimension $n+1$ where $c({\bf \sigma} )$ is left unchanged under the
action of the symmetric group. Within the RS subspace, the occupation
fractions may be conveniently expressed as the moments of a
probability distribution $P(m)$ over the range $-1\le m\le 1$
\cite{MZII}.

\begin{equation}
c  (\sigma _1 ,\sigma _2 ,\ldots ,\sigma _n ) =
\int _{-1} ^1 dm \; P(m) \prod _{a=1} ^n
\left( \frac{1+m \sigma ^a} {2} \right) \qquad .
\end{equation}
$P(m)$ is the distribution of Boolean magnetizations previously
introduced in the paper.

At this stage of the analysis it is possible to perform the analytic
continuation $n \to 0$, since all the functionals have been expressed
in term of the generic number of replicas $n$. Such a process lead to
a self-consistent functional equation for the order parameter $P(m)$. In
the limit of interest $T \to 0$, in order to properly describe the
accumulation of the Boolean magnetization to the border of its domain
($m\in [-1,1]$),  it is convenient to
introduce the rescaled variables $z$, implicitly defined by the relation
$m=\tanh (z/T)$. Calling $R(z)$ the
probability distribution of the $z$s, we obtain
\begin{eqnarray}
R(z) = && \int _{-\infty} ^\infty \frac{du}{2\pi } \cos (u z) \exp
\left\{ - \alpha (1-p) +2 \alpha (1-p) \int _0 ^\infty dz_1 R(z_1)
\cos ( u \min (1,z_1 )) \right. \nonumber \\
&& \left. - \frac 34 \alpha p + 3 \alpha p \int _0 ^\infty dz_1 dz_2
R(z_1) R(z_2 )\cos ( u \min (1,z_1 ,z_2 )) \right\} \qquad .
\label{eqr}
\end{eqnarray}

As discussed in detail in ref. \cite{MZII}, the above type of
equations admit an infinite sequence of rapidly converging
exact solutions of the form
\begin{equation}
R(z) = \sum _{\ell = -\infty}^{\infty} r_\ell \; \delta \left(
z - \frac{\ell}{q} \right) \qquad .
\label{ordre}
\end{equation}
In the above expression, $1/q$
is the resolution of the rescaled variable $z$ which eventually goes to
zero.
Equation (\ref{eqr}), leads to the following set of coupled equations for
the coefficients $r_\ell$'s
\begin{equation}
r_\ell = \int _0 ^{2\pi} \frac{d \theta}{2\pi} \cos (\ell \theta ) \exp
\left( \sum _{j=1} ^q \gamma _j ( \cos (j \theta ) -1 ) \right)
\label{dintyr}
\end{equation}
for all $\ell=0,\ldots ,q-1$ where
\begin{eqnarray}
\gamma _j/ \alpha &=& 2 \;(1-p)\; r_j+3\; p \; r_j
\left(1-r_0 -2 \sum_{\ell=1}^{j-1} r_\ell -r_j\right)  \ , \; \forall j =
1,\ldots ,q-1  \nonumber \\
\gamma _q/\alpha  &=& (1-p) \left(1-r_0-2 \sum_{\ell=1}^{j-1}
r_\ell \right)+
\frac{3}{4}\; p \; \left(1-r_0-2 \sum_{\ell=1}^{j-1} r_\ell
\right) ^2  \qquad .
\label{gammaj}
\end{eqnarray}

By looking for the value of $\alpha$ at which the internal energy
(\ref{expre}) becomes positive, we are able to recover the results
discussed previously in the text.
For $p<p_0$, the transition is continuous and the solution
of the equations up to $\alpha_c$ is simply $r_0=1, r_\ell=0$ ($l=1,...,q$).
At $\alpha_c$, the coefficients $r_\ell$ become continuously positive.
For $p>p_0$, the coefficients $r_\ell$
jump discontinuously to a finite value beyond $\alpha_c$.
It follows that in order to find the point where the discontinuous
transition first takes place, one should look, within the RS scheme, for
the point $p_0$ at which the derivative of the order parameters $r_\ell$ at
$\alpha_c$ diverge.

We may expand the saddle point equations (\ref{dintyr},\ref{gammaj}) to
the second order in the parameters $r_\ell$ and $ s\equiv 1-r_0$.  We find
\begin{eqnarray}
 r_\ell &=& \alpha (1-p) r_\ell+\frac{3}{2}\; \alpha \; p
\; r_\ell \; \left( s -2 \sum_{k=1}^{\ell-1} r_k -r_\ell \right)-
 \alpha^2 \; (1-p)^2 \; r_\ell\; s +
\frac 12 (1-p)^2 \alpha^2
\sum_{j=1}^{\ell-1} r_j r_{\ell-j}+\\ \nonumber
&+& (1-p)^2 \alpha^2 \sum_{j=1}^{q-\ell-1} r_j r_{\ell+j}+
\frac 12 (1-p)^2 \alpha^2 r_{q-\ell} (s-2 \sum_{k=1}^{\ell-1} r_k)
\; \; \; , \; (\ell=1,...,q-1)
\label{tr1}
\end{eqnarray}
and, for $\ell=0$,
\begin{eqnarray}
s &=& \alpha (1-p) s+3 \alpha p \left[
\sum_{j=1}^{q-1} r_j (s-2 \sum_{\ell=1}^{j-1} r_\ell -r_j)+
\frac{1}{4} (s-2 \sum_{\ell=1}^{q-1}  r_\ell)^2 \right]- \\ \nonumber
&-& \alpha^2 (1-p)^2 \left[ \sum_{j=1}^{q-1} r_j^2+
(\frac{s}{2}-\sum_{\ell=1}^{q-1} r_\ell)^2 \right]- \frac 12
\alpha^2 (1-p)^2 s^2
\; \; .
\label{tr2}
\end{eqnarray}
The analysis of the linear terms in the above equations shows that the
threshold is given by
\begin{equation}
\alpha_c (2+ p) = \frac{1}{1-p} \qquad ,
\qquad ( 0\le p < p_0 ) .
\end{equation}
Next, we expand around the latter by
posing $\alpha=\frac{1}{1-p}+x$, $r_\ell =B_\ell x$ and $s=A x$.  At
the critical point $p_0$, the above quantities $\{B_\ell,A\}$ should
diverge in order to have a first order jump when $x \to 0^+$.
We then assume that $B_\ell = \lambda_\ell A$, with $\lambda_\ell = O(1)$
and $A \to \infty$, discarding irrelevant $O(x^2)$ corrections to the
order parameters. We find $q$ equations for $p_0$ and $\lambda_\ell$,
$\ell=1,\ldots , q-1$.
\begin{equation}
0= \frac 34\; \frac{1-2p_0}{1-p_0} - \sum _{j=1}^{q-1} \lambda _j +
\sum _{j=1}^{q-1} \lambda _j ^2 + \left(
\sum _{j=1}^{q-1} \lambda _j  \right) ^2
\label{t1}
\end{equation}
and, for $\ell=1,\ldots , q-1$,
\begin{eqnarray}
0 &=& \frac 32 \frac{p_0}{1-p_0} \lambda _\ell \left( 1 - 2 \sum
_{j=1}^{\ell -1} \lambda _j - \lambda _\ell \right) + \frac 12
\sum _{j=1} ^{\ell -1} \lambda _j \lambda _{\ell - j} +
\nonumber \\ && \sum _{j=1} ^{q -\ell -1} \lambda _j \lambda _{\ell+ j}
+ \lambda _{q-\ell} \left( \frac 12 -
\sum _{j=1}^{q-1} \lambda _j  \right) -\lambda _\ell
\label{t2}
\end{eqnarray}
Though we have not been able to find an exact solution to
(\ref{t1},\ref{t2}), the above equations can be easily solved
iteratively, leading to a value of $p_0 \simeq 0.41$.
A more detailed discussion of the equations for $p_0$ is given
in \cite{MZIII}, in which the connection with the calculation
of ref.~\cite{Achlioptas}
is made explicit by showing that $p_0=2/5$ is
indeed a lower bound for $p_0$.

The exactness of the above results depends on the validity of the RS
assumption introduced in the functional saddle-point equations.
For $p<p_0$, such an assumption turns out to be correct, leading
to a threshold value which as been proven to be exact
also by other methods~\cite{Achlioptas}
For $p>p_0$, the change in the order parameter $P(m)$  (or $R(z)$) at the
threshold becomes discontinuous and the solutions of the RS equations
account only for an upper bound of the true threshold.
The exact solution of the self--consistency equations lies outside the RS
subspace to which we have restricted our analysis.
The exact determination of the SAT/UNSAT threshold for discontinuous
transitions requires the introduction of a more general (and much more
complicated) symmetry breaking
scheme in the equations, the so called Replica Symmetry  Breaking, which
embodies the RS subspace as a particular case.

It is worth noting that in the SAT region $\alpha < \alpha_c(p)$, the RS
theory is believed to be exact and allows for the estimation of quantities of
interest such as the typical number of solutions or the probability
distribution of the variables over all solutions. Some rigorous
probabilistic results are given in \cite{Talagrand}.

\subsection{Comments on the replica approach}

In the previous paragraphs, replicas are introduced as a
trick to compute the average value $\ll \log Z[{\bf C }]\gg$ from the
integer moments $\ll Z[{\bf C }] ^n \gg$ of the generating function (4).
As long as the number of variables $N$ is finite, $a(n) \equiv \ll
Z[{\bf C }] ^n \gg$ is an analytic function of $n$ and
grows less than exponentially at large $n$, $a(n) < (2^N)^n$. Invoking a
theorem of Carlson, $a(n)$ is uniquely known from its values at the
nonnegative integers. Therefore the analytic continuation to real
$n\to 0$ is unambiguous. However, due to the saddle-point calculation
of Section II.B the limit $N \to \infty$ is made first and the
analytic continuation requires the introduction of some additional
hypotheses.

The most natural continuation scheme, called Replica Symmetry (RS) has
recently been shown to be exact at high temperature $T$
\cite{Tala1,Tala2} for the K-SAT model. Though not
explicitly proven in \cite{Tala1,Tala2}, it is reasonable to think that
RS should also hold at $T=0$ for small ratios of clause per variable $\alpha$.
Indeed, in a simple constraint satisfaction model, the
RS hypothesis has been shown \cite{Gardner} to be exact in the range
$0<\alpha <\alpha _c$ giving thus access to the exact value
of the threshold $\alpha _c$ \cite{Cover}. Note that self-consistency
criteria of the calculation of the local stability of the saddle-point
over $c(\vec \sigma )$ found in Section II.B are satisfied by the RS
hypothesis in this range.

What happens when RS fails, {\em e.g.} above (respectively in the
vicinity of) $\alpha _c$ for 2-SAT (resp. 3-SAT) ?
Among all models for which RS fails, the so called Random
Energy Model (REM) \cite{Derrida} has been the only one rigorously solved
so far. Its exact solution can be reproduced \cite{Gross} using
another Replica Symmetry Broken (RSB) scheme designed by Parisi
\cite{MPV}.
Within such a scheme, the analytic continuation is
performed by  an iterative hierarchical procedure characterized by a
closed algebraic structure at each  stage of the hierarchy  \cite{MPV}.
The known random mean-field models (i.e. models
with a complete graph of interaction) appear to be divided in two main
classes. A first one
for which the complete solution requires an infinite iteration of the
Parisi scheme (e.g. the SK model \cite{MPV}) and a second one for which the
first step  already provides the correct result.
In the latter case, the successive steps of the RSB scheme lead to saddle
point equations having as solutions the first step result \cite{Gross}.
It is worth noting here that in the random 3-SAT problem, like
several random mean-field models known to exhibit a discontinuous transition
and be solvable by the one step RSB ansatz, the energy can be expressed as 
a sum of products of up to three spins $S_i$.  
Therefore, the use of the one-step  
RSB hypothesis has promise for
analyzing the SAT/UNSAT transition of 3-SAT.  We expect that there will be 
differences resulting from the fact that the 3-SAT model 
involves a sparse graph of interactions.  This permits heterogeneous orderings
not possible in the mean-field models, for which all degrees of freedom are 
frozen in the ordered phase.

\section{Experimental Tests}
We have run experiments to test the theory between $K=2$ and $K=3$,
finding thresholds and assigning an exponent $\nu$ for the narrowing
of the critical region by finite-size scaling.  The data obtained for
this study is collected and presented in Figs 2a and 2b, which show the fraction
of formulae that are unsatisfiable as a function of $\alpha$ for sample sizes
from $N = 50$ to the largest practical size for each value of $p$.  To obtain the
data in Figs 2a and 2b, we take a sample of formulae with the desired value of $p$,
and for each formula, starting well inside the SAT phase, add clauses until the
formula first becomes unsatisfiable.  The cumulative distribution of the values of
$\alpha$ at which this occurs provides the curves in Figs 2.  From 10,000 to 40,000
formulae were studied for each value of $N$ and $p$ shown.  For cases with $p > p_0$,
our scripts used as their core the TABLEAU implementation of the Davis-Putnam
search algorithm\cite{davis:procedure,TABLEAU}.  For cases with $p < p_0$ a variant called MODOC
was used\cite{MODOC}.  This adds binary resolution to eliminate 2-clauses
by the relation $( p \vee q ) \wedge ( ( \neg q ) \vee r) = p \vee r$.

Finite-size scaling of the critical
region is done by plotting all quantities against the rescaled
variable
\begin{equation}
        y = N^{\frac{1}{\nu}} (\alpha - \alpha_c(K,N))/\alpha_c(K,N) ,
\label{fss}
\end{equation}
which ``opens up'' the critical region in samples with large N so that
data from all sizes collapse onto a single universal curve
\cite{KirkSel}.  If $\alpha_c(K,N)$ were constant, all the curves for a given
value of $K$ would pivot about a single point, $\alpha_c(K,\infty )$.  This occurs
in the finite-size scaling analysis of the random graph ensemble and is a good
approximation at large $K$ for K-SAT\cite{KirkSel}.  But there are significant
additional size dependences present for small values of $K$, as evidenced in
Fig. 3, which shows $K = 2$, ($p = 0.0$) on a fine scale.  The
successive crossings of pairs of curves for increasing values of N provides a
rough measure of
$\alpha_c(K,N)$ (e.g. estimate $\alpha_c(K,50)$ as the point where the fraction UNSAT
for $N = 50$ crosses the fraction UNSAT for $N = 100$).  If we make the ad hoc
assumption that the added size dependence is due to the variation with $N$
of $\alpha_c(K,N)$, then we do not have to specify $\alpha_c(K,N)$ for each $N$.
The data reduction required is to choose values of $\alpha_c(K,\infty ),
\nu$ for which all the transformed curves are parallel, that is, shifted by
$(\alpha_c(K,N) - \alpha_c(K,\infty ))/ \alpha_c(K,\infty )$.
An example of such a reduction is shown in
Fig 4, for the case $p=0.0$.  Using this methodology, we obtained rescaled curves
for all of the data which varied smoothly with $p$, as shown in Fig. 5.

We find (Fig. 6) good
agreement between the observed and predicted values of $\alpha_c$,
with an error which increases slowly from $p=0.41$ to $p=1.0$.  We
also show the bounds obtained by rigorous methods in Fig. 6.  Lower
bounds are obtained by showing that some analyzable algorithms, such
as unit clause propagation\cite{algo} find SAT solutions with a finite
probability\cite{Frieze}.  Upper bounds use the fact that the
probability of finding a satisfying assignment is bounded by the
expected number of solutions.  Refined versions of this
argument\cite{Dubois,Kirousis} partially eliminate the
high degeneracy of some solutions\cite{Kamath}.  Both methods have
been applied to $(2+p)$--SAT in \cite{Achlioptas}, yielding the dashed
and dotted lines plotted in Fig. 6.

In figure 7, we show values of $\nu$ obtained from finite-size scaling
analysis described above.  Below $p_0$, the exponent $\nu$ is
roughly constant and equal to 2.8, the value found for 2SAT.  This indicates
that the critical behavior along the second order transition line in Fig. 6
is dominated by the 2-clauses in the formulae.  When we include additional
corrections to scaling in $y(\alpha,N)$ and in the probability of UNSAT,
following the classic prescription\cite{Fisher}, we find that $\nu$
may be as large as 3, the value that occurs in the percolation transition
for undirected random graphs\cite{Bollobas}.  The UNSAT phase for
$K=2$ is one in which "constraint loops" become so ubiquitous that almost
certainly there is some literal that implies its converse.  It is likely that
the 2SAT transition results from percolation of these loops, and
is in the same universality class as random graph
percolation, differing only in the corrections to its scaling behavior.

Above $p_0$, $\nu$ drops rapidly to 1.5.  For $K \gg 3$, the values of
$\nu$ tend to 1.0, a result which can be understood in the
``annealed'' limit discussed in \cite{MZII,Kamath}.

It is surprising that finite size scaling holds in the presence of
discontinuous behavior of the order parameter characterizing the
backbone. But this discontinuity is accompanied by smooth behavior of
other thermodynamic quantities, e.g.  entropy, as first discussed by 
Gross and M\'ezard in \cite{Gross}.  
First order
transitions in pure solids involve two (or a finite number of) phases
and do not exhibit critical fluctuations or scaling laws with
non-integer exponents.  The random first-order
transition taking place for $p>p_0$, into
an infinite number of distinct ground states, displays features of
both first and second orders. This mixed behaviour has also recently been
observed in random-field models \cite{Sourlas}.

Previous work showed that the cost of running the best heuristics,
depth-first search algorithms with backtracking \cite{davis:procedure}
increased exponentially with $N$ for any value of $\alpha$ for $K =
3$, with a prefactor that could be mapped into a universal function by
plotting it as a function of $y$\cite{SelKirk}.  The cost was
maximized at $\alpha_{.5}(K,N)$, so we have obtained cost data at this
value of $\alpha$ for $p = 0., .2, .4$ and $.6$ over the range of N
that could be searched.  The plot in Fig. 8a shows that the median
cost increases linearly with N for $p < p_0$.  It increases
dramatically over a smaller range of $N$ for $p > p_0$.  Fig. 8b
confirms that this increase is exponential already for $p = 0.6$.

Discontinuous nucleation of UNSAT regions due to the breakdown of
replica symmetry and the ``backbone'' of frozen spins conveniently
explain the apparent inevitably high cost of heuristic search near the
phase boundary.  Heuristics which proceed by "asserting" the possible
value of a spin make early mistakes by mis-asserting a backbone spin,
and take a long time to backtrack to correct their mistakes.  Even if
the backbone can be identified before the depth-first search begins,
the problem that remains is one of organizing the search over the
remaining spins which lie on the boundaries of "nuclei" or partial
solutions to find the lowest energy arrangements of the whole
solution, also involving much wasted effort to explore an exponential
subspace.

The experiments, shown in Figs. 9a ($K = 2$) and 9b ($K = 3$),
confirm that the appearance of
the backbone is discontinuous for $K = 3$, and support the prediction of a continuous
appearance of the backbone for $K = 2$.
Above the threshold, the fraction of frozen
spins found in small samples by exhaustive enumeration to locate all
ground states is relatively insensitive to N.
At and below the threshold, the fraction of frozen spins decreases
rapidly with increasing N.  While the samples which could be studied
are too small to permit extrapolation, the results are consistent with
$<f( K,\alpha )>$ vanishing below $\alpha_c$.  

The existence of a ``backbone'' has previously been reported in the
traveling salesman problem\cite{KirTou}, with only a few bonds
differing in many near-optimal tours.  This observation has recently
been turned to advantage by heuristics\cite{Morgen} which identify the
backbone links and concentrate attention on the small subproblems
which remain.  This may prove to be a generally valid approach.
Efficient means of finding the backbone will be specific to each
problem type, but should nonetheless provide a step ahead in algorithm
efficiency.
Moreover, many worst-case NP-complete problems occurring
in practice contain a mix of tractable and intractable constraints.
Our results suggest that search algorithms that exploit as much of
the tractable structure as possible may in fact scale polynomially in the
typical case. In much of the work on search methods in, for example, Artificial
Intelligence and Operations Research, one already informally follows the
methodology of exploiting tractable problem structure in worst-case
intractable problems. However, our hybrid model provides the first formal
explanation why such a methodology can work so well in practice.
Below a certain threshold fraction of intractable constraints, the overall
behavior is dominated by the tractable substructure of the problem, leading
to an overall efficient, polynomial time solution method.


\newpage
\begin{figure*}[]
\centerline{\vbox{\hbox{
\psfig{figure=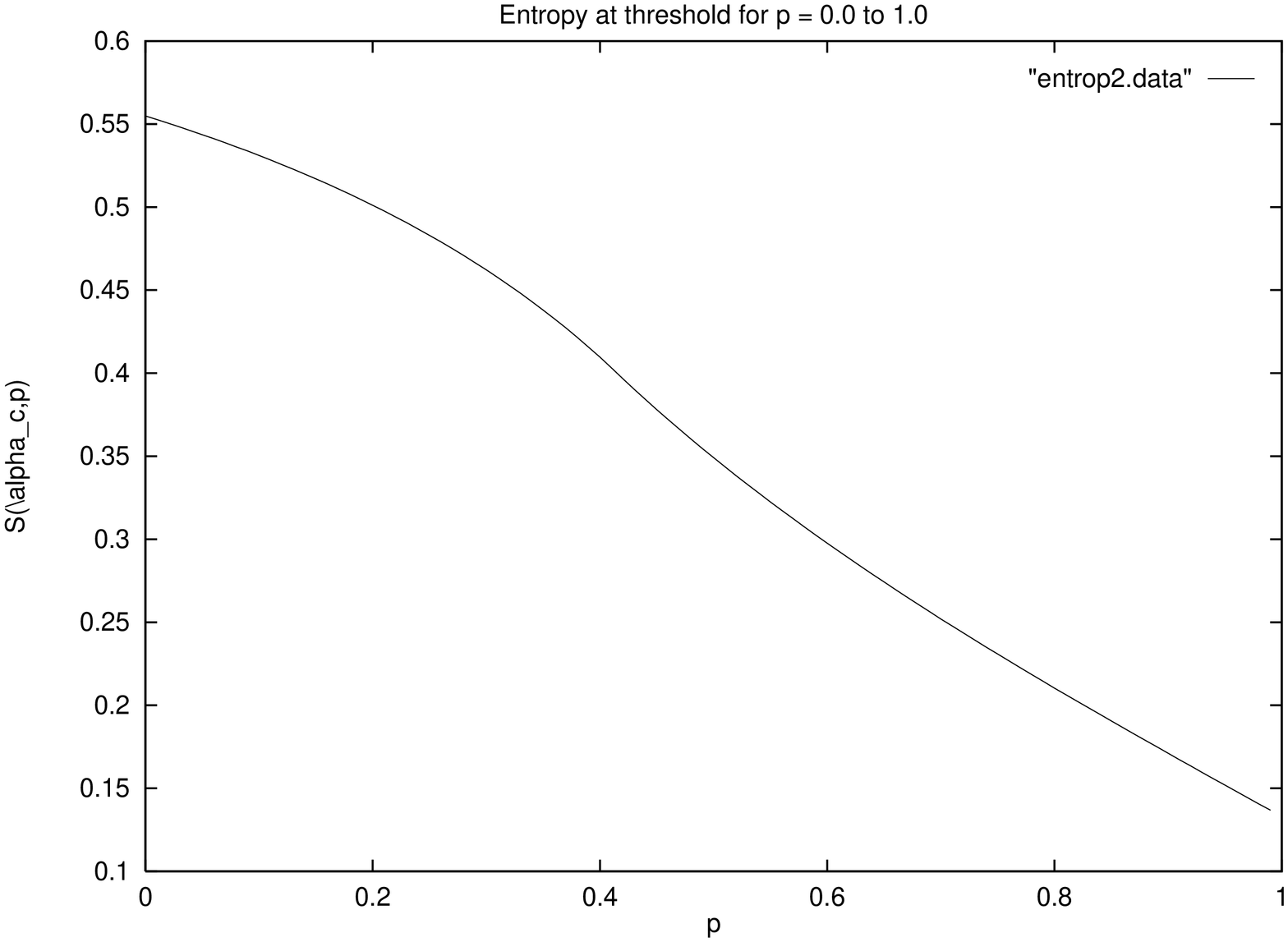,width=3in,height=3in,angle=-90}
}}}
\vspace{0.15in}
1.  Ground state entropy at $\alpha_c(p)$ versus $p$, predicted by the RS
theory of \cite{MZII,MZIII}.  For $p < p_0$, $\alpha_c(p) = 1/(1-p)$.  For
$p > p_0$, we have used the estimates of $\alpha_c(p)$ obtained by
finite-size scaling.
\end{figure*}

\begin{figure*}[]
\centerline{\vbox{\hbox{
\psfig{figure=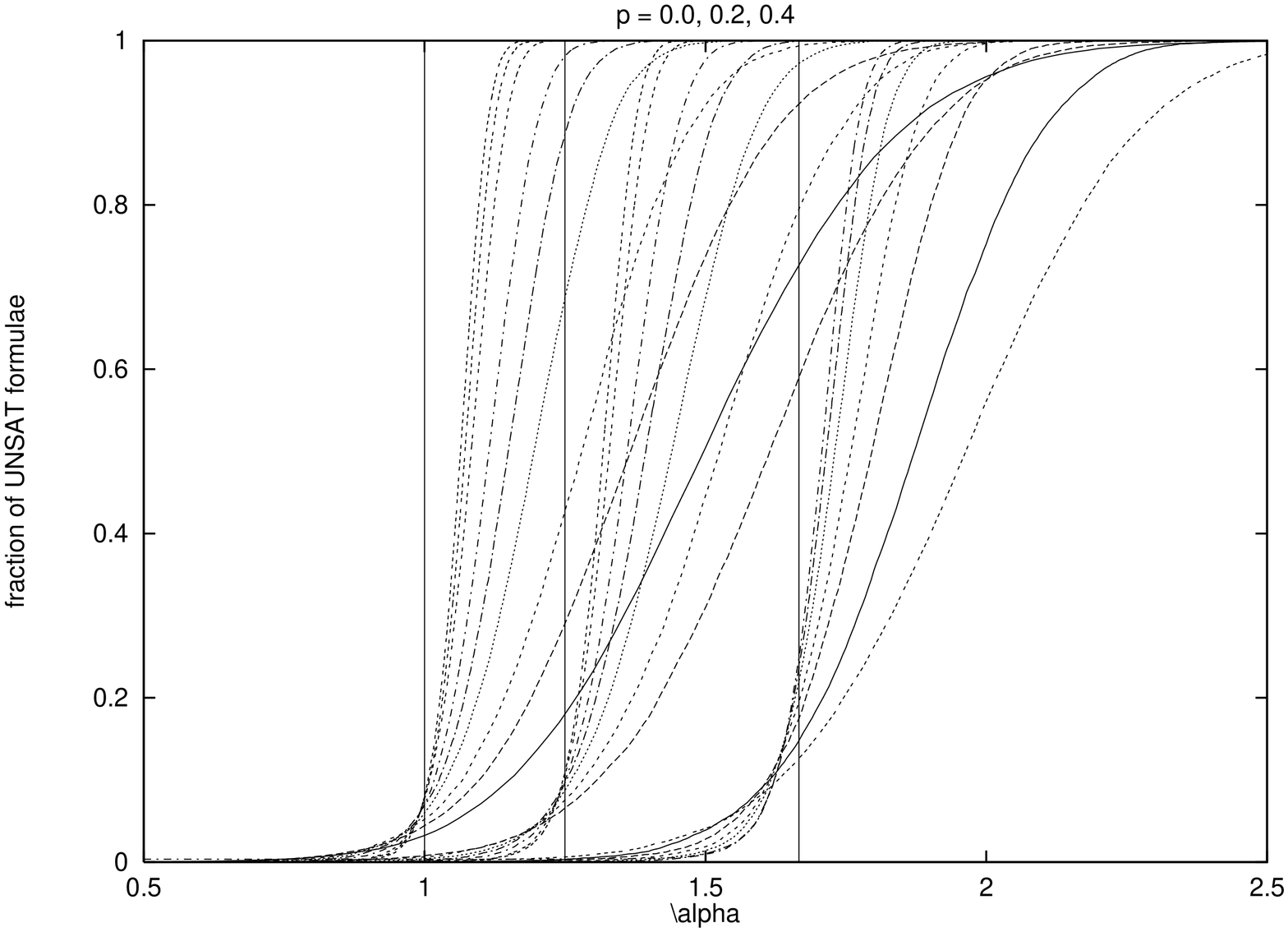,width=5in,height=3in,angle=-90}
}}}
\vspace{0.15in}
2a.  Raw data used in this study for $p=0.0$, 0.2, and 0.4.
Vertical lines mark the thresholds for each
value of $p$.  For $p=0.0 (K=2)$, data are plotted for $N =$ 50, 100, 200,
500, 1000, 2000, 5000, 7500, and 10000.  For $p=0.2$, values of $N$ are
100, 200, 500, 1000, 2000, 5000, and 7500.  For $p=0.4$, values of $N$ are 100,
200, 500, 1000, 2000, 3500, and 5000.
\end{figure*}

\newpage
\begin{figure*}[]
\centerline{\vbox{\hbox{
\psfig{figure=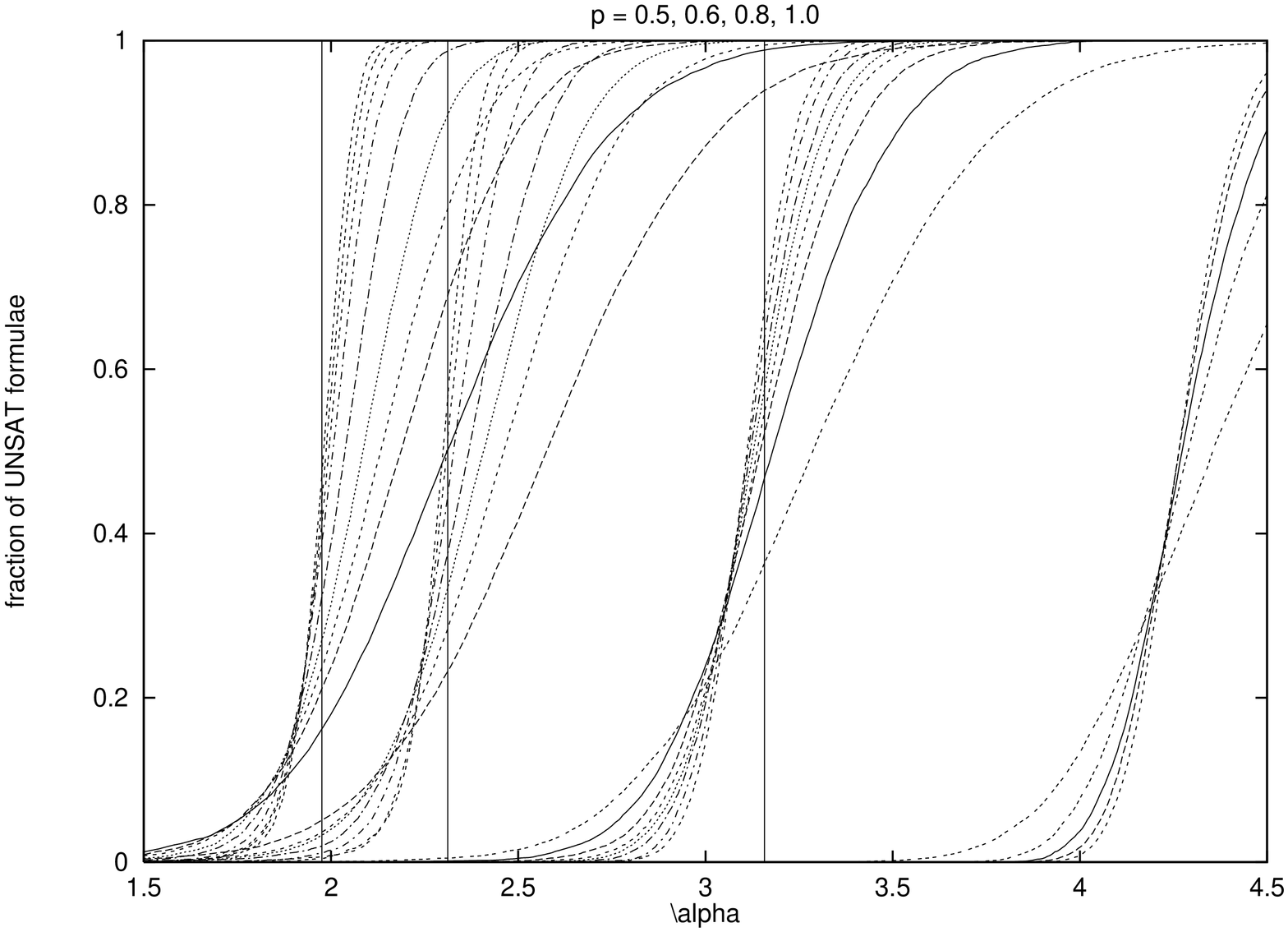,width=5in,height=3in,angle=-90}
}}}
\vspace{0.15in}
2b.  Raw data for $p=0.5$, 0.6, 0.8 and 1.0.  Thresholds marked are determined from the
RS theory (an overestimate).  For $p=0.5$, values of $N$ are 50, 100, 150, 250, 500,
1000, 1500, 2000, and 2500.  For $p=0.6$, values of $N$ are 50, 100, 150, 250, 500,
1000, and 1500.  For $p=0.8$, values of $N$ are 50, 100, 150, 200, 250, 300, 400, and
500.  For $p=1.0 (K=3)$, values of $N$ are 50, 100, 150, 200, and 250.
\end{figure*}

\begin{figure*}[]
\centerline{\vbox{\hbox{
\psfig{figure=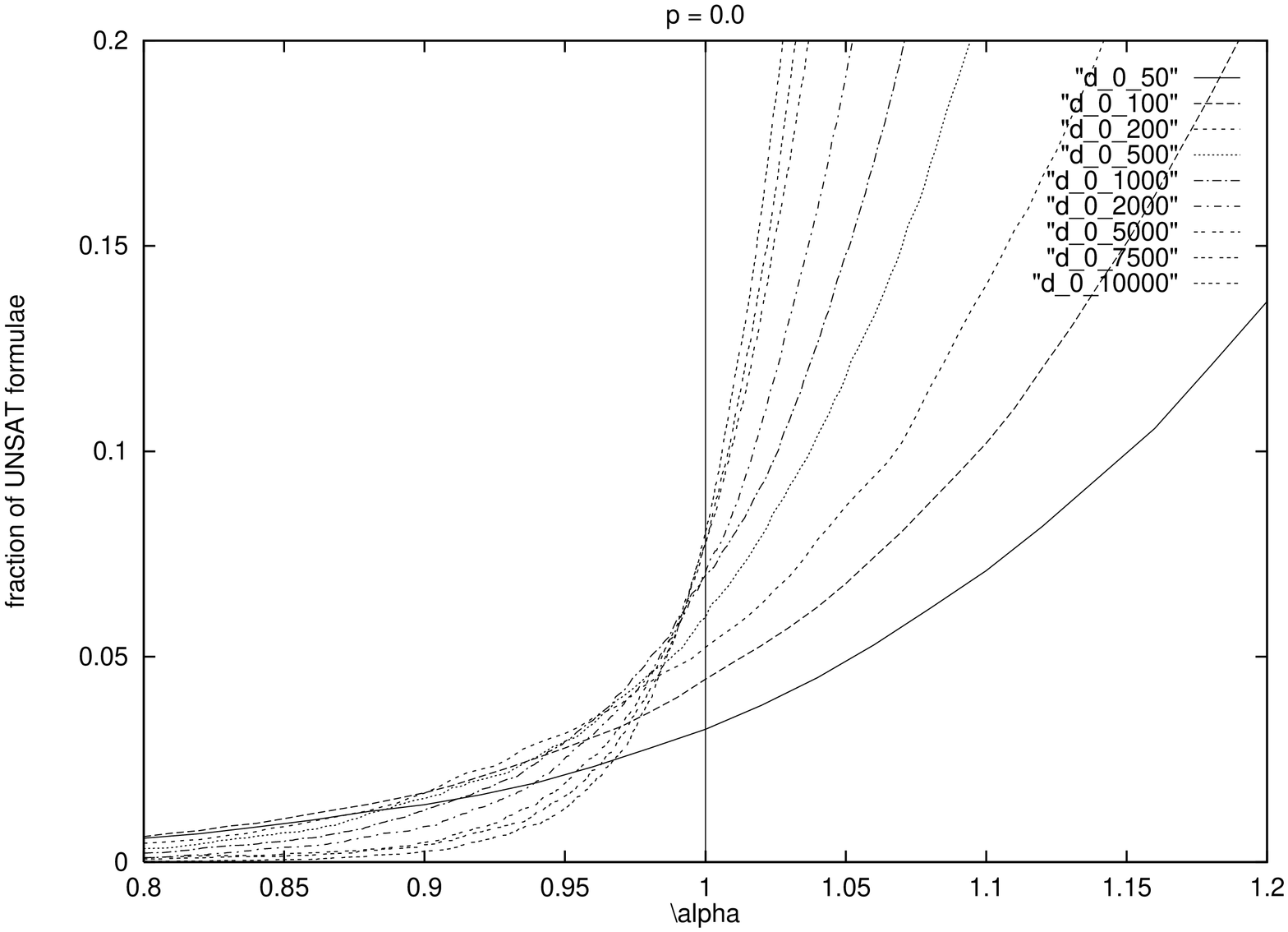,width=5in,height=3in,angle=-90}
}}}
\vspace{0.15in}
3.  Plot of data obtained for $K=2 (p = 0.0)$, shown on a finer scale to exhibit
size dependence.
\end{figure*}

\newpage
\begin{figure*}[]
\centerline{\vbox{\hbox{
\psfig{figure=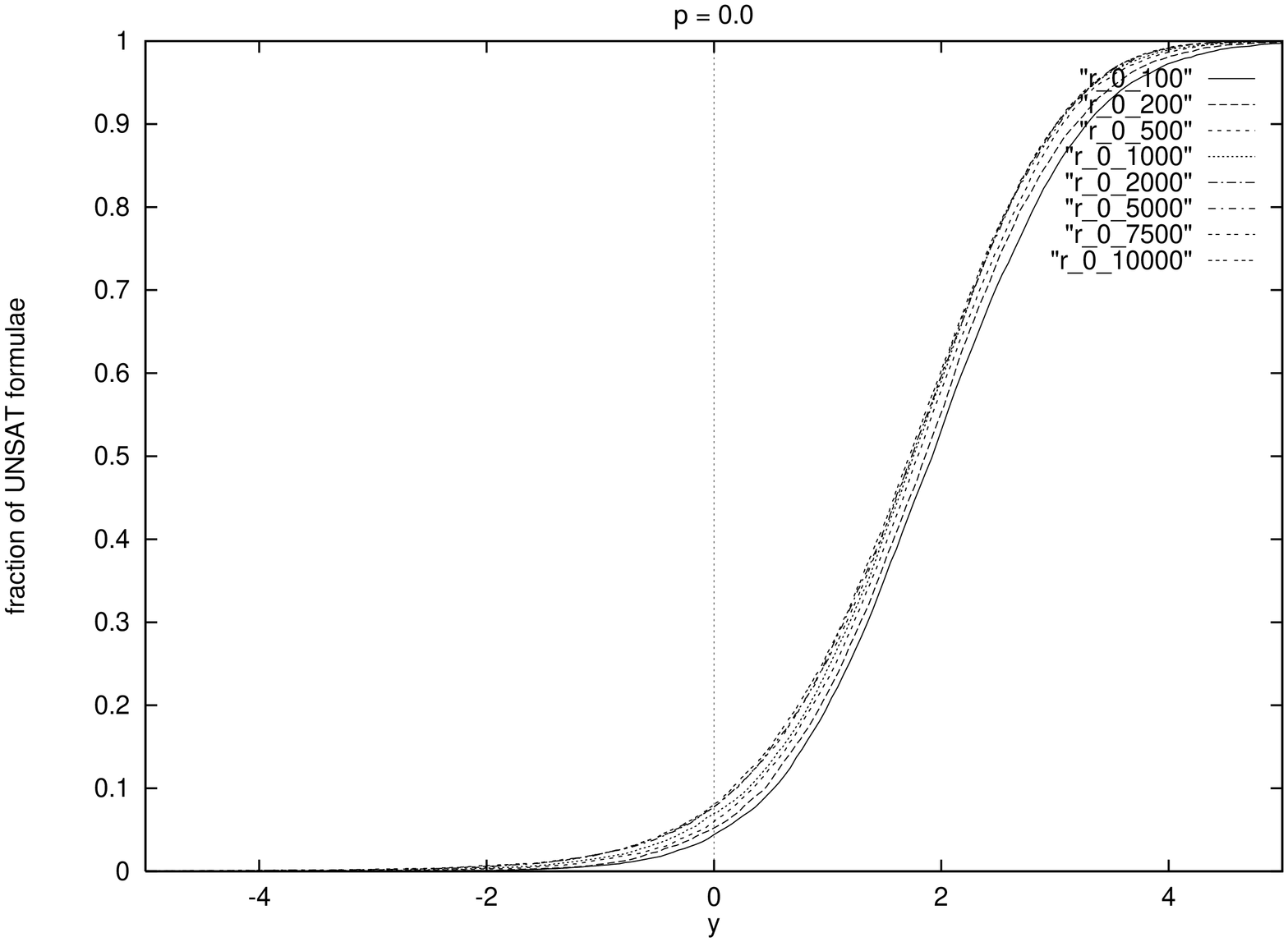,width=5in,height=3in,angle=-90}
}}}
\vspace{0.15in}
4. Rescaled data for $K = 2 (p = 0.0)$ using $\nu = 2.8$.  The curves for
different values of $N$ are parallel, converge to a limit for large $N$, and
are monotonic in $N$ at each value of $y$.
\end{figure*}

\begin{figure*}[]
\centerline{\vbox{\hbox{
\psfig{figure=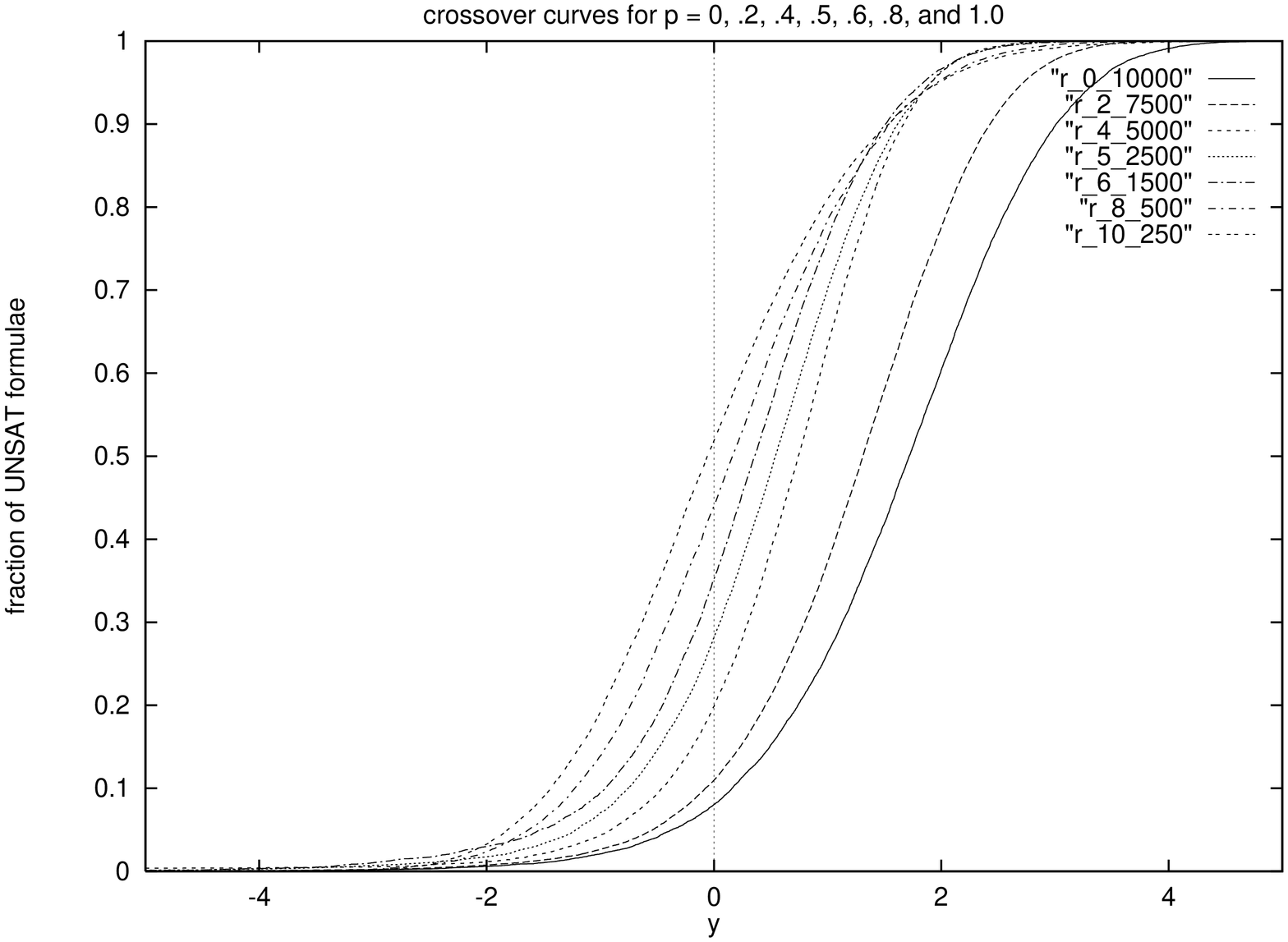,width=5in,height=3in,angle=-90}
}}}
\vspace{0.15in}
5.  Rescaled data for all $p$, using the largest values of N which could be obtained,
with $alpha_c$ and $\nu$ determined as described in text.
\end{figure*}

\newpage
\begin{figure*}[]
\centerline{\vbox{\hbox{
\psfig{figure=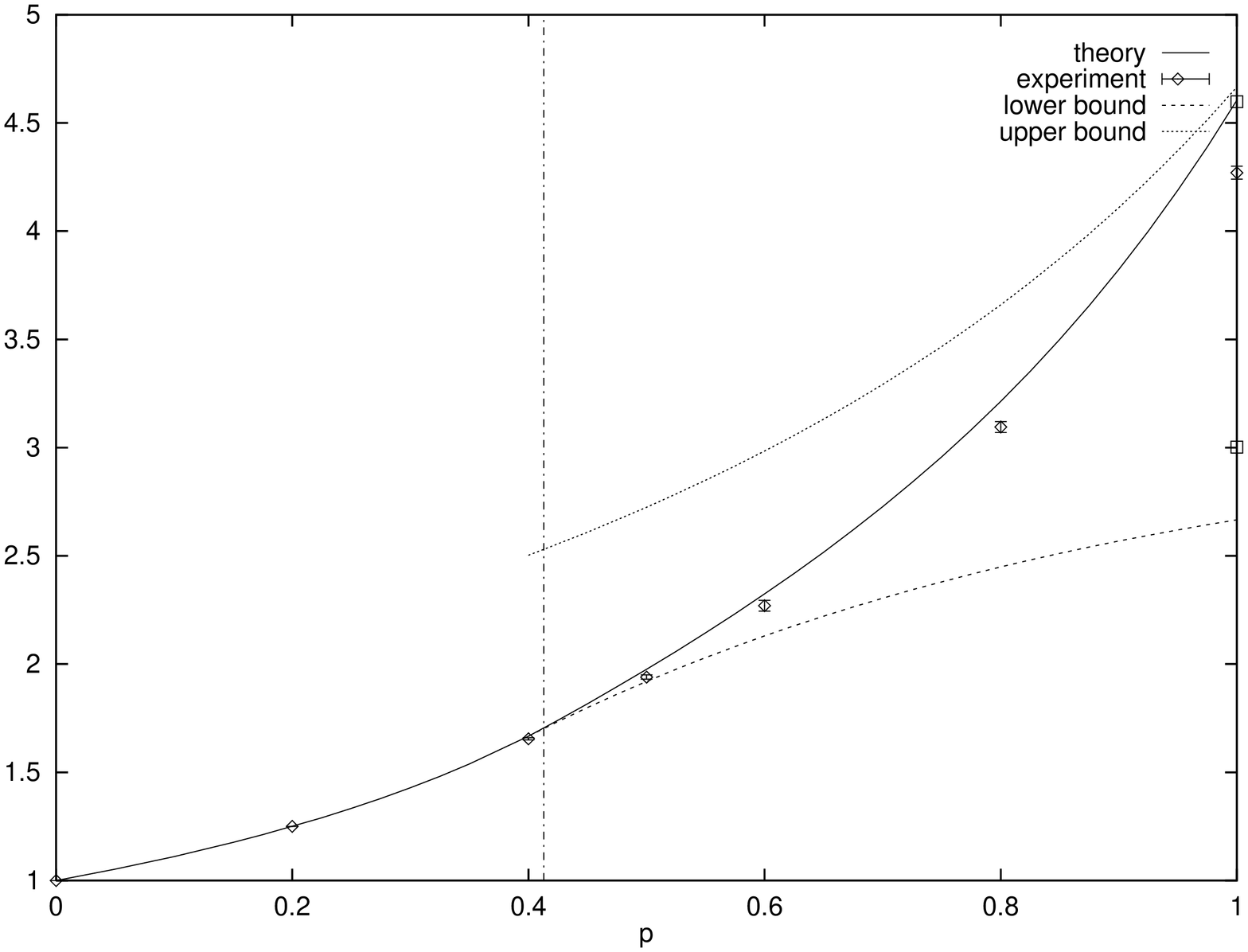,width=5in,height=3in,angle=-90}
}}}
6.  Theoretical and experimental results for the SAT/UNSAT transition
in the 2+p-SAT model.  The vertical line at
$p_0$ separates the continuous from the discontinuous transition.  The
full line is the replica-symmetric theory's predicted transition,
believed exact for $p<p_0$, and the diamond data points with error bars are
results of computer experiment and finite-size scaling.  The other two
lines show upper and lower bounds obtained in \cite{Achlioptas}, while
the stronger  upper bound due to \cite{Kirousis}, and the best known
lower bound, due to \cite{Frieze}, are indicated by square data points.
\end{figure*}

\begin{figure*}[]
\centerline{\vbox{\hbox{
\psfig{figure=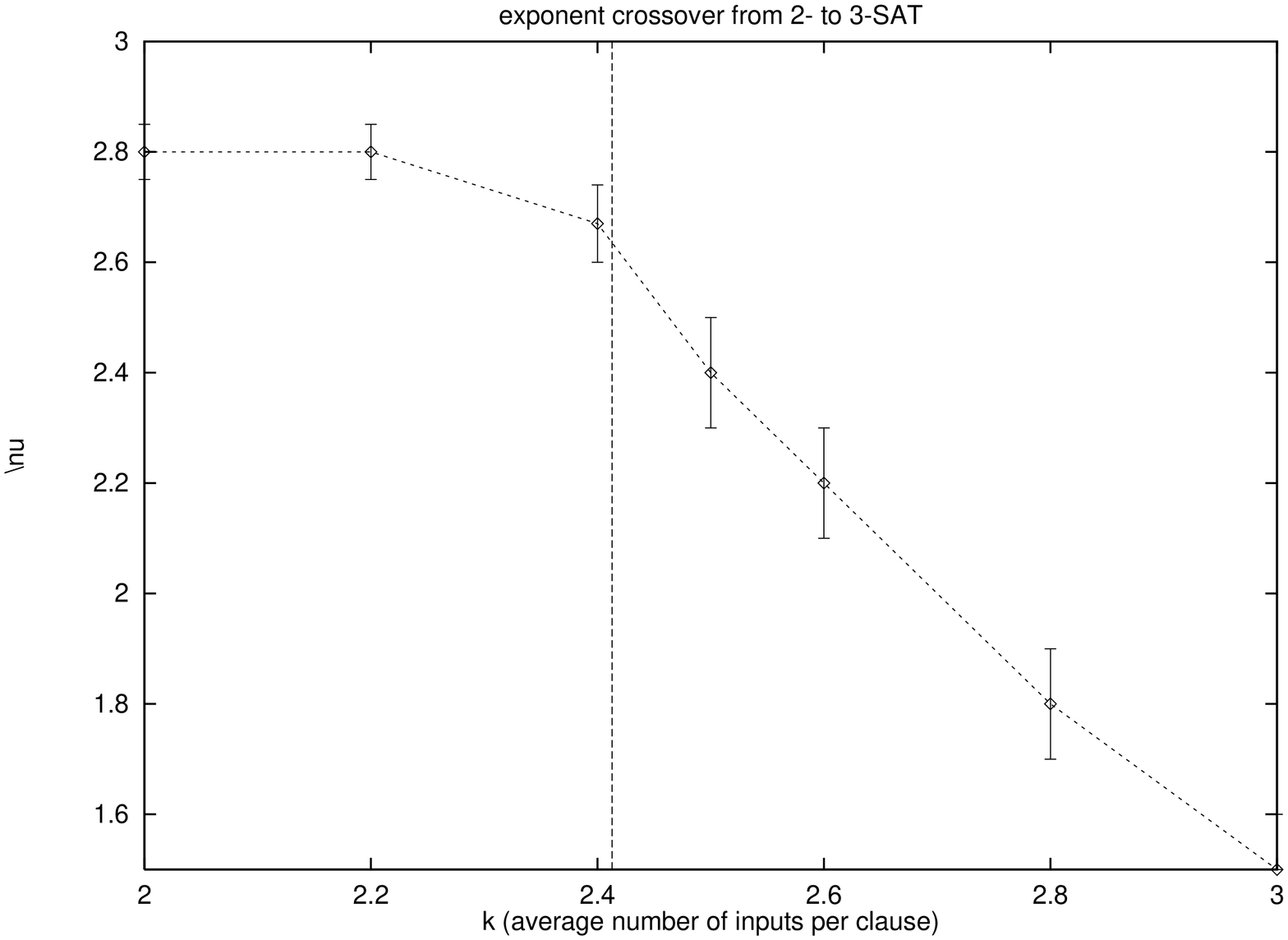,width=5in,height=3in,angle=-90}
}}}
7.  Crossover seen in the exponent $\nu$ governing the width of the
critical regime, as $K$ increases from 2 to 3.
\end{figure*}

\newpage
\begin{figure*}[]
\centerline{\vbox{\hbox{
\psfig{figure=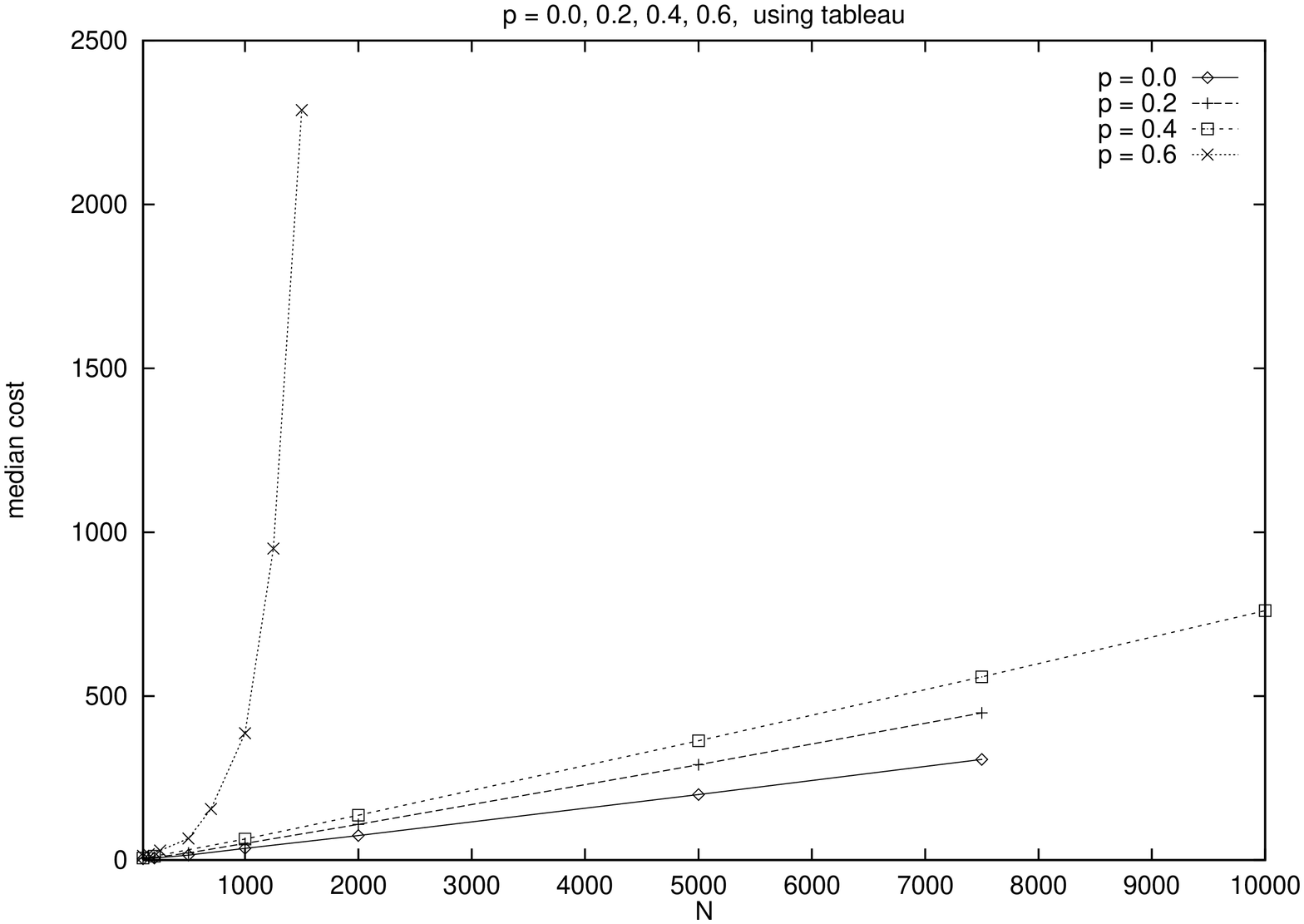,width=5in,height=3in,angle=-90}
}}}
\center{Fig.\ 8a: Median computational cost, linear scale.}
\vspace{0.15in}

\centerline{\vbox{\hbox{
\psfig{figure=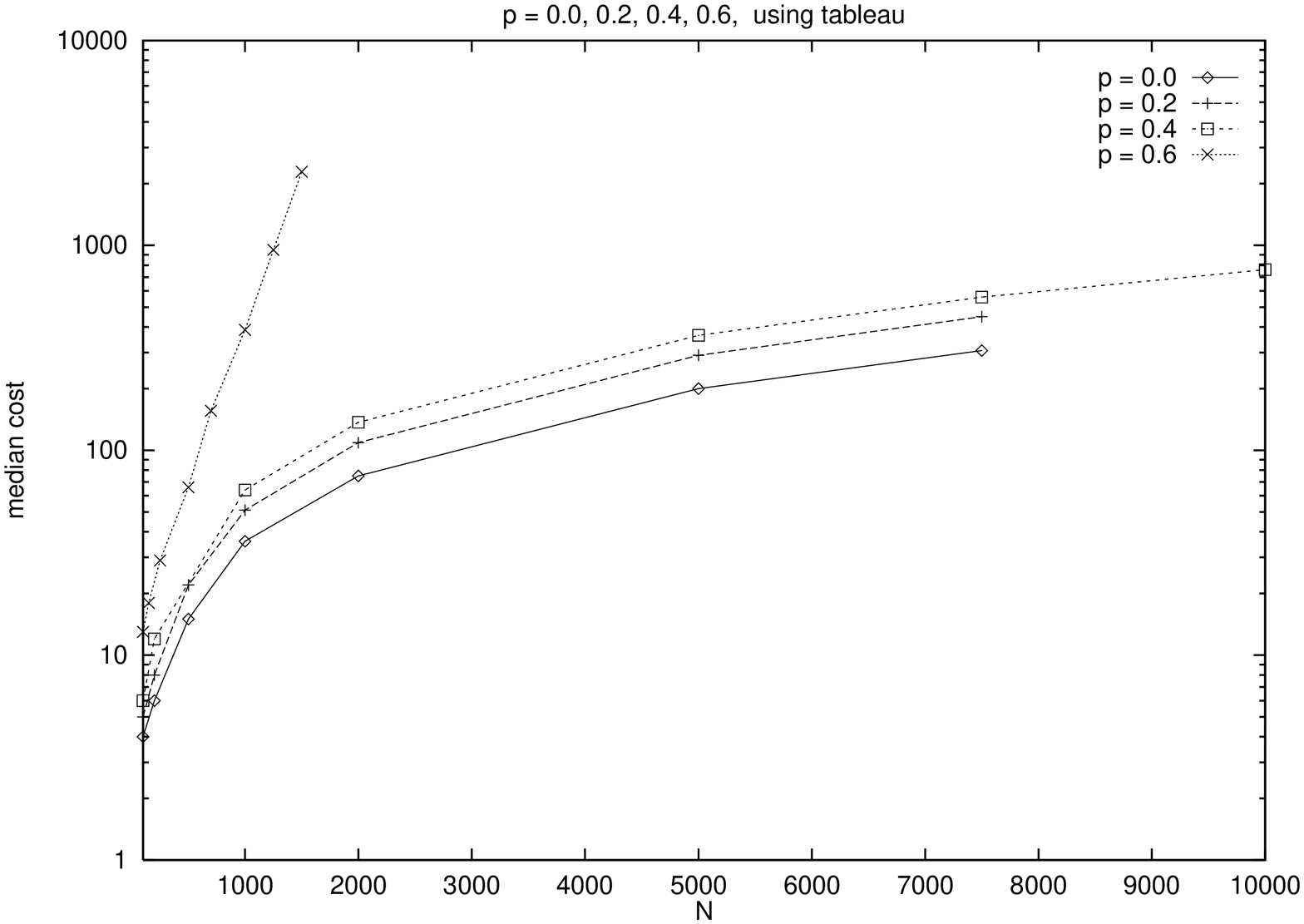,width=5in,height=3in,angle=-90}
}}}
\center{Fig.\ 8b: Median cost, semilog scale.}
\vspace{0.15in}

8.  Median computational cost of proving a formula SAT or UNSAT using
the TABLEAU search method, for $p$ ranging from 0 to 1.  The data in (a)
is plotted on a linear scale, appropriate for
the cases with $p < p_0$.  The semi-log plots in (b) show an
exponential dependence of cost on $N$ for $p > p_0$.
\end{figure*}

\newpage

\begin{figure*}[]
\centerline{\vbox{\hbox{
\psfig{figure=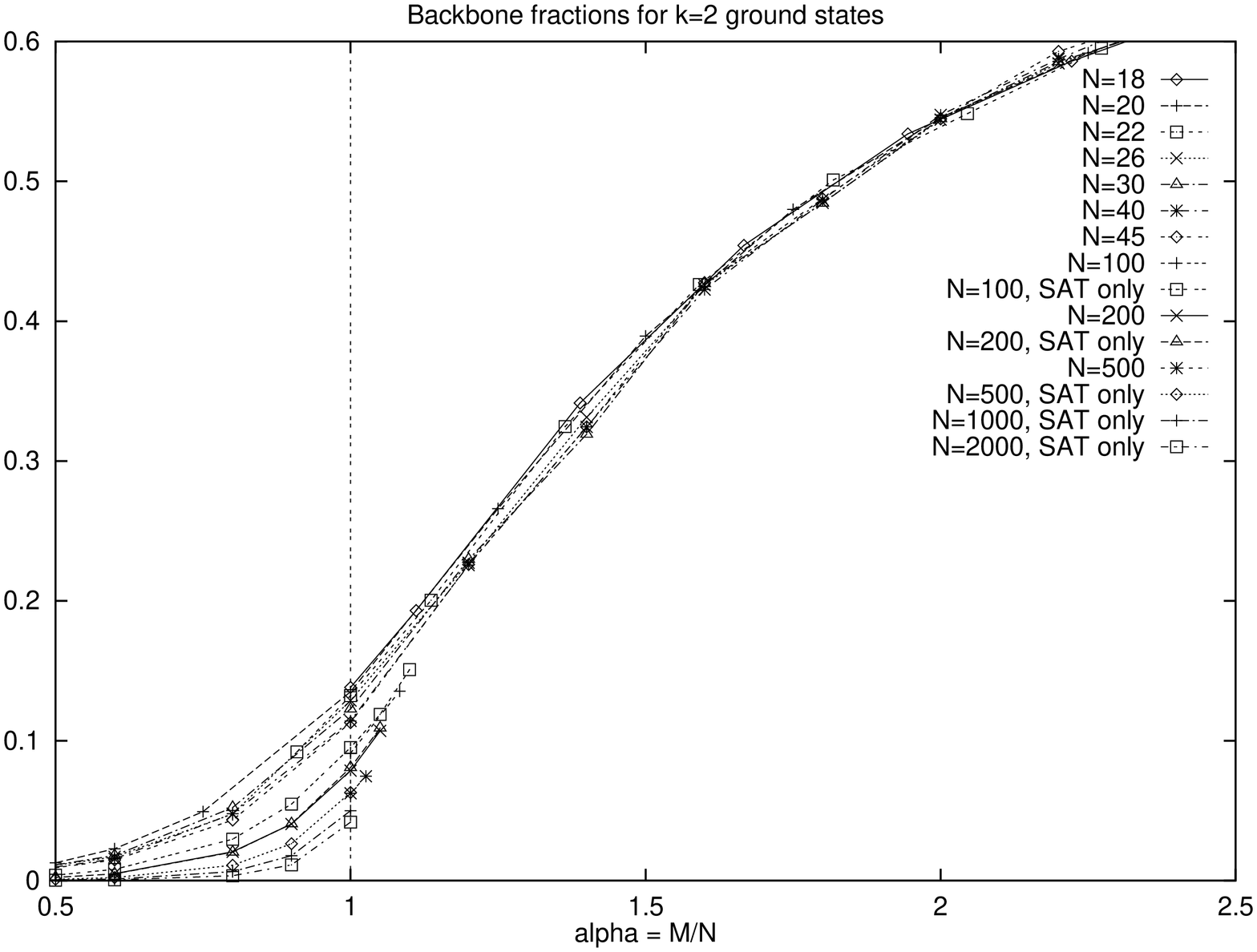,width=5in,height=3in,angle=-90}
}}}
\center{fig.\ 9a: Backbone fraction for $K = 2$.}
\vspace{0.15in}

\centerline{\vbox{\hbox{
\psfig{figure=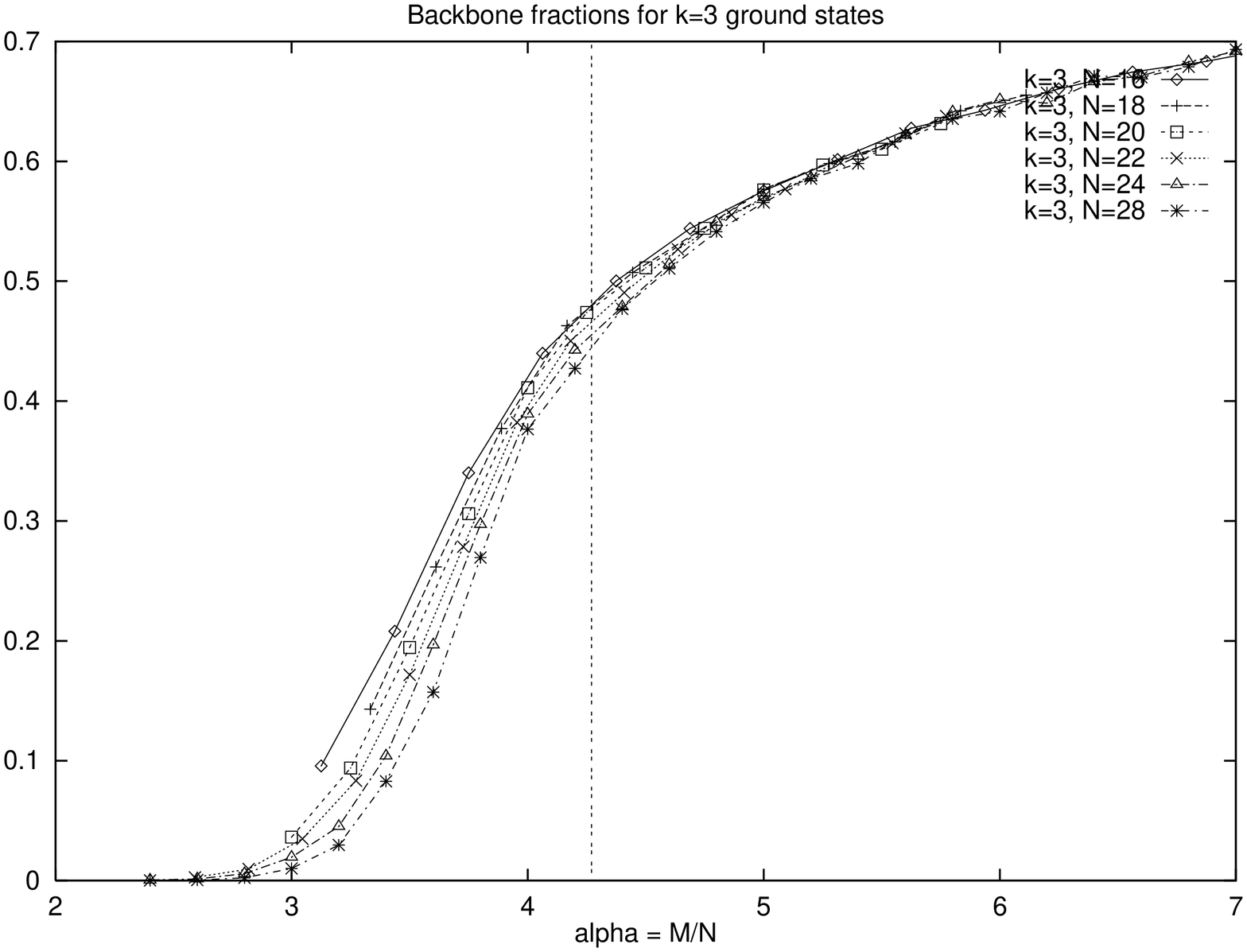,width=5in,height=3in,angle=-90}
}}}
\center{fig.\ 9b: Backbone fraction for $K = 3$.}
\vspace{0.15in}

9.  Backbone fractions as a function of $\alpha$ averaged over many
samples for $K=2$ (a) cases with $N =$ 18 to 45
and $K=3$ (b) cases with $N =$ 18 to 28.  The vertical
lines mark the SAT/UNSAT thresholds in the limit $N \to \infty$.
For 2-SAT, data obtained from larger sizes $N=100, 200, 500$ show that the
backbone fraction at the threshold decreases to zero.
\end{figure*}


\begin{thebibliography}{99}
\bibitem[1]{rm} Email: monasson@lpt.ens.fr
\bibitem[2]{rz} Email: zecchina@ictp.trieste.it
\bibitem[3]{sk} Email: kirk@watson.ibm.com
\bibitem[4]{bs} Email: selman@cs.cornell.edu
\bibitem[5]{lt} Email: lidrort@cs.huji.ac.il

\bibitem{Hayes}
        for an introductory discussion see {\sc Hayes}, B., {\em
        American Scientist} {\bf 85}, 108 (1996)

\bibitem{FuAnd}
        {\sc Fu}, Y.-T., and {\sc Anderson}, P. W.,
        in {\sl Lectures in the Sciences of Complexity}, D. Stein (ed.),
        (Addison-Wesley, 1989), p. 815.

\bibitem{Fisher}
	   Scaling and critical phenomena have a vast literature.
         For a masterful summary and review, see
	   {\sc Fisher}, Michael E., {\sl Reviews of Modern Physics} {\bf 70},
         pp. 653-681 (1998).

\bibitem{exact}
        {\sc Goerdt}, A., in {\sl Proc. 7th Int. Symp. on Mathematical
        Foundations of Computer Science}, 264 (1992);
        {\sl Journal of Computer and System Sciences}, {\bf 53}, 469 (1996) \\
        {\sc Chv\`atal}, V., and {\sc Reed},B., in
        {\sl Proc. 33rd IEEE Symp. on Foundations of Computer
        Science}, 620 (1992)

\bibitem{Cook}
        {\sc Cook},S.A., {\it The complexity of theorem--proving procedures
        }, in {\sl Proc. 3rd Ann. ACM Symp. on Theory of Computing},
        Assoc. Comput. Mach., New York, 151 (1971)

\bibitem{NPC}
        {\sc Garey}, M., and {\sc Johnson}, D.S.,
        {\sl Computers and Intractability;
        A guide to the theory of NP--completeness}, W.H. Freeman and Co., San
        Francisco, 1979; C. Papadimitriou, {\sl Computational Complexity},
        Addison--Wesley, 1994;

\bibitem{algo}
        {\sc Aspvall}, B, {\sc Plass}, M.F. and {\sc
        Tarjan},R.E. {\em Inf. Process. Lett.}{\bf 8}, 121 (1979)

\bibitem{KirkSel}
        {\sc Kirkpatrick}, S., and {\sc Selman}, B.,
        {\em Science} {\bf 264}, 1297 (1994)

\bibitem{NIPS}
        {\sc Kirkpatrick}, S.,
        {\sc Gyorgyi}, G., {\sc Tishby}, N., and {\sc Troyansky}, L.,
        NIPS Proceedings Vol 6, 439-446 (1993).

\bibitem{Friedgut}
        {\sc Friedgut}, E.,
        preprint, presented at DIMACS 97.

\bibitem{note1} Above $\alpha_c$ proving UNSAT is still hard, with a lower
       bound on search cost that is exponential in $N$ with a coefficient
       that decreases as a power law in $\alpha$. See
       ``On the Complexity of Unsatisfiability Proofs for Random k-CNF
       Formulas'', P. {\sc Baume}, R. {\sc Karp}, T. {\sc Pitassi} and
       M. {\sc Saks}, {\em Proc. STOC-98}, 1998 561-571.

\bibitem{mitchell:hard}
        {\sc Mitchell}, D., {\sc Selman}, B., and {\sc Levesque},H.,
        ``Hard and Easy Distributions of SAT problems,''
        {\em Proc.\ of Am. Assoc. for Artif. Intell. AAAI-92} (1992), 456-465.

\bibitem{AI}
        Issue 1--2, {\em Artificial Intelligence} {\bf 81}, {\sc
        Hogg}, T., {\sc Huberman}, B.A., and {\sc Williams}, C.,
         Eds., (1996)


\bibitem{SelKirk}
        {\sc Selman}, B., and {\sc Kirkpatrick}, S.,
        {\em Artificial Intelligence} {\bf 81}, 273 (1996)

\bibitem{SimAnn}
        {\sc Kirkpatrick}, S., {\sc Gelatt} Jr., C.D.,
        and {\sc Vecchi}, M.P.,
        {\em Science} {\bf 220}, 339 (1983)

\bibitem{kro}
        {\sc Hertz}, J., {\sc Krogh}, A., and {\sc Palmer}, R.G.,
        {\em Introduction to the theory of neural computation},
        Addison-Wesley, Redwood City (CA), 1991

\bibitem{MPV}
        {\sc M\'ezard}, M., {\sc Parisi}, G., {\sc Virasoro} {M.A.},
        {\em Spin Glass Theory and Beyond}, World Scientific, Singapore, 1987

\bibitem{MZI}
        {\sc Monasson}, R., and {\sc Zecchina} R.,
        {\em Phys. Rev. Lett.} {\bf 76}, 3881 (1996)

\bibitem{MZII}
        {\sc Monasson}, R., and {\sc Zecchina} R.,
        {\em Phys. Rev.} {\bf E 56}, 1357 (1997)

\bibitem{MZIII}
        {\sc Monasson}, R., and {\sc Zecchina} R.,
        {\em J. Phys. A: Math. Gen.} {\bf 31}, 9209 (1998)

\bibitem{physcomp96}
        {\sc Monasson}, R., and {\sc Zecchina} R.,
        {\sc Kirkpatrick}, S., {\sc Selmann}, B., {\sc Troyansky}, L.,
        ``Phase Transition and Search Cost in the $2+p$ SAT Problem'',
        proceedings of {\em PhysComp96}, {\sc Toffoli}, T.,
        {\sc Biafore}, M., {\sc Le$\tilde{a}$o}, J., eds., Boston (1996)

\bibitem{Achlioptas}
        {\sc Achlioptas}, D., {\sc Kirousis}, L., {\sc Kranakis}, E.,
        and {\sc Krizanc}, D.,
        ``Rigorous Results for Random $(2+p)$--SAT'',
        {\em submitted to Phys.\ Rev.\ E}.  A preliminary version
        appeared in RALCOM 97.

\bibitem{selfave}
        {\sc Broder}, A.Z., {\sc Frieze}, A.M., and {\sc Upfal}, E.,
        {\em Proc. 4th Annual ACM--SIAM Symp. on Discrete Algorithms},
        322 (1993)

\bibitem{eilat}
        {\sc Monasson, R.},
        {\em Phil. Mag.} {\bf B 77}, 1515 (1998);
        {\em J. Phys.} {\bf A 31}, 513 (1998)

\bibitem{Talagrand}
        {\sc Talagrand}, M.,
        Huge Random Structures and mean field models for spin glasses,
        Doc.\ Math.\ J.\ DMV Extra Volume ICM I, 507 (1998)

\bibitem{Tala1} {\sc Talagrand}, M. {\em Probability and Related Fields}
        {\bf 110}, 109-176 (1998).

\bibitem{Tala2} {\sc Talagrand}, M. ``The high temperature case of the
         K-sat problem''. To appear in {\em Probability Theory and Related
         Fields}

\bibitem{Gardner} {\sc Gardner}, E., {\sc Derrida}, B. {\em J. Phys. A}
         {\bf 22}, 1983 (1989)


\bibitem{Cover} {\sc Cover}, T.M. {\em IEEE Trans. Electron. Comput.}
        {\bf EC-14}, 326 (1965)


\bibitem{Derrida} {\sc Derrida}, B. {\em Phys. Rev. B} {\bf 24},  2613 (1981)


\bibitem{Gross} {\sc Gross}, D., {\sc M\'ezard}, M. {\em Nuclear Phys. B}
        {\bf 240}, 431 (1984)


\bibitem{davis:procedure}
        {\sc Davis}, M., and {\sc Putnam}, H.,
        ``A computing procedure for quantification theory,''
        {\em J.\ Assoc.\ Comput.\ Mach.}, {\bf 7} (1960), 201--215.

\bibitem{TABLEAU}
	{\sc Crawford}, J.M. and {\sc Auton}, L.D. (1993).
  	``Experimental results on the cross-over point in satisfiability problems.''
  	{\em Proc. AAAI-93}, Washington, DC, 21--27.

\bibitem{MODOC}
	{\sc Van Gelder}, Allen (1999).
       "Autarky pruning in propositional model elimination
       reduces failure redundancy"
       Journal of Automated Reasoning (to appear).

\bibitem{Frieze}
        {\sc Frieze}, A., and {\sc Suen}, S.
        ``Analysis of two simple heuristics on a random instance of
        K--SAT'',
        {\em Journal of Algorithms 20} (1996), 312-335.

\bibitem{Dubois}
        {\sc Dubois}, O., and {\sc Boufkhad}, Y.,
        ``A general upper bound for the satisfiability threshold of
        random K--SAT formulas,''
        {\em Journal of Algorithms 24}, 395-420 (1997)

\bibitem{Kirousis}
        {\sc Kirousis}, L., {\sc Kranakis}, E., and {\sc Krizanc},
        D.,
        ``Approximating the unsatisfiability threshold of random
        formulas,''
        {\em Proceedings of the 4th European Symposium on Algorithms},
        (1992), 27-38.

\bibitem{Kamath}
        {\sc Kamath}, A., {\sc Motwani}, R., {\sc Palem}, K.,
        and {\sc Spirakis}, P.,
        ``Tail bounds for occupancy and the satisfiability threshold
        conjecture,''
        {\em Random Structures and Algorithms} {\bf 7}, 59 (1995)

\bibitem{Bollobas}
	{\sc Bollob\'as}, B.,
	``Random Graphs'', (Academic Press, London, 1985).


\bibitem{Sourlas}
        {\sc Angl\`es d'Auriac}, J.-C. and {\sc Sourlas}, N.,
        ``The $3d$ random field Ising model at zero temperature,''
        {\em Europhysics Lett. 39}, (1997) 473-478.


\bibitem{KirTou}
        {\sc Kirkpatrick}, S. and {\sc Toulouse}, G.,
        ``Configuration space analysis of travelling salesman
        problems,''
        {\em Journale de Physique} {\bf 46}, (1985) 1277-1292.

\bibitem{Morgen}
       {\sc Schneider}, J., {\sc Froschhammer}, C., {\sc Morgenstern}, I.,
	{\sc Husslein}, T.,  {\sc Singer}, J. M.,
	``Searching for Backbones -- an efficient parallel algorithm for
	the travelling salesman problem'',
	{\em Computer Physics Communications 96} (1996) 173-188



\end{thebibliography}
\end{document}